# A B<sup>link</sup> Tree method and latch protocol for synchronous node deletion in a high concurrency environment

Karl Malbrain
malbrain@cal.berkeley.edu

## Introduction

A new B<sup>link</sup> Tree latching method and protocol simplifies implementation of Lehman and Yao's high concurrency B<sup>link</sup> Tree together with Jaluta's balanced B<sup>link</sup> Tree methods and Lanin and Shasha's node deletion methods for B<sup>link</sup> Trees. In a high availability NoSQL environment, implementations usually call for periodic global tree reorganizations that can have a significant impact on performance when they occur.   The bottom up approach presented here synchronously removes empty nodes from the tree as they are created by deletion.

The top down methods presented by Jaluta are problematic. As an update search proceeds down the B<sup>link</sup> Tree, a "U" latch to each child node is obtained while the parent node "U" latch is held. Any delays in obtaining the child's "U" latch will curtail concurrent update searches through the parent since "U" latches are incompatible with one another. During node splits and consolidations, an "X" latch is held on the child node while the new upper fence values are posted into the parent node, shutting down all concurrent access to the node.

## New latch modes

Instead of Jaluta's hierarchical "S-U-X" B<sup>link</sup> Tree latch modes that each protect a node exclusively, three completely independent node latches are used for concurrency control. These latches are implemented in a multi-threaded environment by memory fenced latches manipulated by the `__sync_fetch_and_xxx` function family:

- The first latch is a sharable `AccessIntent` and exclusive `NodeDelete` used to drain readers and updaters from accessing an empty node before its removal from the tree. `AccessIntent` is obtained during `Search` using latch-coupling from the parent node to the child node and never blocks. `NodeDelete` is obtained as the final stage before returning the node a free list of available nodes. It is requested only after all pointers to the node have been removed from the B<sup>link</sup> Tree and blocks until all `AccessIntent` latches have drained away.

- The second latch is a traditional, sharable `ReadLock` and exclusive `WriteLock` for the node. These latches are obtained during `Search` using latch-coupling from `AccessIntent`.

- The third latch is a `ParentModification` exclusive latch which is held while a node's fence value is either updated or removed from its parent node during a `Split` or `Consolidation`. It is obtained with latch-coupling from the node's `WriteLock`. This ensures that structure modifications are serialized with respect to one another, while allowing release of the node's `WriteLock` for concurrency during the fence key postings in the node's parents.

The cost of guaranteeing that nodes will not be referenced by concurrent tree operations after they are deleted is the requirement to obtain two node latches at each level of the tree after the root access during `Search`, instead of just one. This expense is mitigated by the absence of latch upgrades or downgrades during `Insert` or `Delete,` and by the high performance of the page latch implementation.

## Latch Compatibility Matrix

A "y" (yes) entry means that two latches of the types specified by the row and column labels for that entry can be simultaneously held on a data item by two different B[link] Tree operations, i.e., the two latch types do not conflict. An "n" (no) entry means that the two latch types cannot be concurrently held on a data item by distinct operations, i.e., the latch types conflict. Note that after `NodeDelete` is requested, no further `AccessIntent` requests are possible since all references to it have previously been removed from the tree and this is marked N/A. The table indicates the latch type requested across the X axis, and the existing latch(es) previously granted down the Y axis.

```
                        Latch Requested

                    AI    ND    RL    WL    PM
E   AccessIntent    Y     N     Y     Y     Y
X   NodeDelete      N/A   N/A   Y     Y     Y
I   ReadLock        Y     Y     Y     N     Y
S   WriteLock       Y     Y     N     N     Y
T   ParentModification Y  Y     Y     Y     N
```

## Implementation Details

There are five latch modes for each node organized into three independent latches. Each latch consists of a thirty-two bit word which is manipulated by atomic instructions to implement a sharable/exclusive latch:

- (latch 1) `AccessIntent`: The read latch on the parent node will be released, and the child node `ReadLock` or `WriteLock` will be requested next while holding `AccessIntent`. This mode will never block since the `NodeDelete` that is incompatible will not be requested until all references to the node have been removed from the tree.

- (latch 1) `NodeDelete`: Pointers to the node have been removed from the tree under a `WriteLock`, making the node inaccessible. Wait for all `AccessIntent` latches to drain away.

- (latch 2) `ReadLock`: Read the node. Incompatible with `WriteLock`.

- (latch 2) `WriteLock`: Allow modification of the node. Incompatible with `ReadLock` and other `WriteLocks`.

- (latch 3) `ParentModification`: Serialize the change of the node's upper fence values in the parent nodes. Incompatible with another `ParentModification`, but compatible with the other four latches on the node.

## Tree Operations

The following methods are implemented in the associated sample C code:

`Search`: While descending the B[link] Tree from the root to the target node at the requested tree level, a `ReadLock` on a parent is latch-coupled with obtaining `AccessIntent` on the child node, which never blocks since `NodeDelete` is obtained only after all references have been removed from the B[link] Tree. `AccessIntent` is latch-coupled with obtaining `ReadLock` or `WriteLock` as appropriate on the child node. Requests for key values beyond the node's upper fence value are directed to the node's right sibling, which is processed with the same `AccessIntent` and `ReadLock` and upper fence key check. Accessing an empty node undergoing deletion results in a sibling traversal using the link field to the node to its left.

`Update/Insert`: A key value is added or updated to a node under the `WriteLock` obtained during `Search` for the root/branch/leaf node being posted and the `WriteLock` is released. The same `Update/Insert` module is used for root, branch, and leaf nodes. If the new key doesn't fit in the node because it is full, the node is `Split` first and the `Update/Insert` is restarted.

`Split`: A `ParentModification` latch is obtained for the node over the existing `WriteLock`. Note that this will block if a previous `Split/Consolidation` for this node is still underway. The upper half of the contents are split into a new right sibling node (which is latched with a `ParentModification`), and the `WriteLock` on the left node is released. The new median fence value for the left node is `Inserted` into the appropriate parent, and the new right sibling node's upper fence key is `Updated` in its parent node to switch it from the left node to the new right node. Note that different parent nodes for the left and right nodes are possible if the parent `Splits`. The `ParentModification` latches on the left and right nodes are released.

`Delete`: A key is removed from a node under a `WriteLock` obtained during `Search` by setting its key delete bit. The `WriteLock` is released if the node is not empty, otherwise the node is `Consolidated` with its existing right sibling, if any. If an upper level fence key was deleted, correct the grandparent's node keys. If the root page was reduced to a single child, collapse the root node by replacing it with the child node's contents.

`Consolidation`: A `WriteLock` is obtained for the right sibling node, and its contents (including its right sibling link field) replace the empty left node's contents under the existing `WriteLock`. The right sibling node's deleted bit is set, and the right node's link field is set to point to the left node, which will direct subsequent searches to the consolidated node. A `ParentModification` latch is obtained for the consolidated node. Note that this blocks if another `ParentModification` is already underway due to concurrent B[link] Tree operations. The `WriteLock` for the left node and the `WriteLock` on the right node are released allowing further updates by concurrent tree operations. The old left node upper fence key is `Deleted` from its parent, and the old right node upper fence key is `Updated` in its parent to point to the left node. Note that these upper fence keys might be in different parent nodes. The `ParentModification` latch on the left node is released. A `NodeDelete` latch is obtained for the deleted right node which waits for all existing `AccessIntent` latches to drain, and a `WriteLock` is obtained for the right node which waits for all `ReadLock` or `WriteLock` to drain.. There are now no pointers and no other latches for the right node, and it is placed onto a free-list for use in subsequent splits. The `NodeDelete` and `WriteLock` latches are released from the now free node.

The structure modifications to the upper tree levels are serialized by the `ParentModification` latch while the node remains available for subsequent `Searches`, `Update/Inserts`, and `Deletes` while the structure modification proceeds.

## Deleting Fence Keys

When deleting a child's upper fence key from a parent which is also the parent's upper fence key,

two courses of action are possible. This occurs whenever a `Consolidated` node and its right sibling happen to have different parent nodes.

    1) The entire child node delete can be ignored, leaving the empty child node in the B$^{link}$ Tree for later use by `Inserts`. This is the approach taken by Jaluta, Lomet, and others, and leaves the key ranges intact for the parent nodes. It suffers from a lack of symmetry with `Insert`, and shuts down node `Deletes` at the grandparent level where nodes are never emptied. The grandparent's key space partitions are never changed to reflect the current contents of the B$^{link}$ tree. Periodic tree reorganizations are typically scheduled to remove empty nodes from the tree.

    2) The change in the parent's fence key can be propagated up the B$^{link}$ Tree to the grandparent level. This approach will permanently narrow the key range of the parent node, pushing any and all re-inserted keys under the parent's right sibling. The sample code takes this approach.

## Latch Coupling

At several points in the implementation, latch coupling is used to acquire a new latch on a node before releasing another. During `Splits/Consolidations` this overlap is extended to include additional node processing before the previous latch is released so that two latches might be held simultaneously. During `Search`, a parent node will remain fully available during any delay in obtaining a child `ReadLock/WriteLock` over `AccessIntent`. This guarantees that the operations conducted under the `ParentModification` latch remain deadlock free.

## Source Code URL

Full implementation in the C language for several variants of multi-process or multi-threaded implementations is available at `https://github.com/malbrain/Btree-source-code` and consists of about 2500 lines of code and comments. It will compile under either unix/linux or WIN32.

## Key Size and Node Layout

The node size is specifiable in bits, from 9 (512 bytes) thru 20 (1 MB) and stores variable sized keys, each from 1 thru 255 bytes in length. Each node has a tree level value (0 for leaf), a delete bit which redirects searches to the link (left) node, a count of slots in use and a count of non-deleted keys.

    Each key in the node has a deleted bit which allows keeping the upper fence key value on the page even after it's deleted. The actual key bytes are allocated space at the top of the node. The space from deleted keys is reclaimed during node cleanup which is done as needed on `Insert` before deciding to `Split`.

    An array of keys and their values is kept sorted in the lower part of each node after its header, with the highest occupied array slot reserved for the node's upper key fence value, even if it is deleted. Each node also has a link field which points to a sibling node. An active node points to its right sibling if overflowing the upper fence key, while a deleted node has its link changed to point to the `Consolidated` node on the left which is used under guidance of the node's delete bit. The link field is updated during `Split` and `Consolidation`.

    Page zero is reserved for new and reused node allocation. Page one is the location of the tree

root. Page two is always the first page of leaves. B$^{link}$ Tree page numbers and Key Value row-id's are all 48 bit quantities.

## Buffer Pool Manager

A Buffer Pool Manager is included in the source code that utilizes the underlying Operating System file mapping interface to map B$^{link}$ Tree segments to program virtual memory. Up to 65536 segments are managed (this is a Linux and WIN32 system limitation). The Buffer Manager will also run without a pool or file mapping, using standard file system reads and writes to access nodes. The number of consecutive nodes per segment is configurable.

## Acknowledgements

Paul Pedersen conducted a line by line code review with the author and ported the code to C++ for inclusion in the MongoDb project.